\documentclass[%
superscriptaddress,
 amsmath,amssymb,
 aps,
]{revtex4-2}

\usepackage{hyperref}
\usepackage{graphicx}
\usepackage{dcolumn}
\usepackage{bm}
\usepackage{amsmath}

\begin{document}

\preprint{APS/123-QED}

\title{Bayesian inference of ion velocity distribution function from laser-induced fluorescence spectra}

\author{S. Tokuda}
\email{s.tokuda.a96@m.kyushu-u.ac.jp}
\affiliation{Research Institute for Information Technology, Kyushu University, Kasuga 816-8580, Japan}

\author{Y. Kawachi}
\affiliation{Interdisciplinary Graduate School of Engineering Sciences, Kyushu University, Kasuga 816-8580, Japan}

\author{M. Sasaki}
\affiliation{College of Industrial Technology, Nihon University, Narashino 275-8575, Japan}

\author{H. Arakawa}
\affiliation{Institute of Science and Engineering, Shimane University, Matsue 690-8504, Japan}

\author{K. Yamasaki}
\affiliation{Graduate School of Advanced Science and Engineering, Hiroshima University, Hiroshima 730-8511, Japan}

\author{K. Terasaka}
\affiliation{Interdisciplinary Graduate School of Engineering Sciences, Kyushu University, Kasuga 816-8580, Japan}

\author{S. Inagaki}
\email{inagaki@riam.kyushu-u.ac.jp}
\affiliation{Research Center for Plasma Turbulence, Kyushu University, Kasuga 816-8580, Japan}
\affiliation{Research Institution for Applied Mechanics, Kyushu University, Kasuga 816-8580, Japan}

\begin{abstract}
The velocity distribution function is a statistical description that connects particle kinetics and macroscopic parameters in many-body systems. Laser-induced fluorescence (LIF) spectroscopy is utilized to measure the local velocity distribution function in spatially inhomogeneous plasmas.  However, the analytic form of such a function for the system of interest is not always clear under the intricate factors in non-equilibrium states. Here, we propose a novel approach to select the valid form of the velocity distribution function based on Bayesian statistics. We formulate the Bayesian inference of ion velocity distribution function and apply it to LIF spectra locally observed at several positions in a linear magnetized plasma. We demonstrate evaluating the spatial inhomogeneity by verifying each analytic form of the local velocity distribution function. Our approach is widely applicable to experimentally establish the velocity distribution function in plasmas and fluids, including gases and liquids.
\end{abstract}

\flushbottom
\maketitle
%
%
\thispagestyle{empty}

\section*{Introduction}
The Maxwellian distribution is the most fundamental velocity distribution function, which holds in thermodynamic equilibrium states. However, it does not necessarily hold for non-equilibrium states. The ions in magnetized plasmas, which are usually far from equilibrium, are subject to other types of the velocity distribution function. The bi-Maxwellian distribution represents the anisotropy of flow and temperature in parallel and perpendicular directions to the external magnetic field \cite{goldston2020introduction}. The ring
velocity distribution is a model of the energetic ion population that is formed immediately after neutral beam injection \cite{moseev2019bi}. The velocity
distribution function of $\alpha$-particles has a fat tail originating from collisions of these particles with the thermal ions and electrons in the background plasma \cite{goldston2020introduction, spitzer2006physics, stix1972heating}. 

Laser-induced fluorescence (LIF) spectroscopy is utilized to observe the velocity distribution of ions or neutral particles in low-temperature plasmas \cite{severn1998argon,  boivin2003laser, claire2006laser}. 
Local features in inhomogeneous plasmas, such as number density, flow velocity, and local temperature, are obtained by fitting the representation of the velocity distribution function in the absorption frequency domain to an observed spectrum.
However, the explicit expression of such a function is often unclear or ambiguous. While intricate factors deform the velocity distribution in non-equilibrium states, as highlighted above, it is not always easy to tell which is dominant/negligible.
Furthermore, changes in internal degrees of freedom, such as the Zeeman and Stark effects, make it more challenging to identify the explicit expression of the velocity distribution function. They deform the shape of the observed spectrum even if the velocity distribution itself is unchanged \cite{boivin2003laser, kelly2016ari, arakawa2019ion}. 
It is crucial to experimentally establish the analytic form of the velocity distribution function and its parameters without ambiguity for understanding the system of interest.

In this study, we propose a novel approach to evaluate which form of the velocity distribution function assumed for an observed spectrum is valid. Our approach is based on the Bayesian model selection \cite{efron1973stein, akaike1980likelihood, mackay1992bayesian, bishop2006pattern}, which quantify the uncertainty in each analytic form of the function assumed for observed data. We formulate the Bayesian inference of the velocity distribution function from an observed spectrum. 
To demonstrate the validity of our approach, we apply our approach to LIF spectra locally observed at some positions in a magnetized helicon plasma. We successfully verify that each local ion velocity distribution function is Maxwellian rather than other candidates.

\section*{Results}
\textbf{Models.}
We introduce our models while briefly reviewing the relationship between the velocity distribution function and their representation in the absorption frequency domain.
To observe the velocity distribution function, laser spectroscopy including LIF utilizes the (non-relativistic) Doppler shift from eigenfrequency $\omega_{0}$ of target particles to absorption frequency $\omega$:
\begin{align}
\omega = \omega_{0} \left( 1+ \frac{v + \Delta v}{c}\right),
\label{eq:Doppler}
\end{align}
where $v + \Delta v$ is the particle velocity parallel to the propagation direction of the incident light, and $c$ is the speed of light.
Note that the second-order Doppler effect is ignored \cite{demtroder1973laser}.
Here, $\Delta v$ is the contribution from the bulk fluid flow, and $v$ is the relative velocity subject to the velocity distribution function defined in the coordinate moving at $\Delta v$.
The frequency shift resulting from the bulk fluid flow is defined as  $\Delta \omega := \omega_{0} \Delta v / c$ for convenience.

If the system is a magnetized plasma in the partial local thermodynamic equilibrium state, $v$ of the ions is subject to the Maxwellian distribution
\begin{align}
f(v; n_{i}, T_{i}):= n_{i} \sqrt{\frac{m}{2 \pi k_{B} T_{i}}} \exp \left ( -\frac{mv^2}{2 k_{B} T_{i}} \right),
\label{eq:Maxwell}
\end{align}
where $n_{i}$, $T_{i}$, $m$ and $k_{B}$ are respectively the ion density, the ion temperature, the ion mass and the Boltzmann constant.
Then we obtain the spectral line shape
\begin{align}
f(\omega; n_{i}, T_{i}, \Delta \omega, \omega_{0}) := n_{i} \sqrt{\frac{mc^2}{2 \pi k_{B} T_{i} \omega_{0}^2}} \exp \left ( -\frac{mc^2(\omega -\Delta \omega -\omega_{0} )^2}{2 k_{B} T_{i} \omega_{0}^2} \right),
\end{align}
where $f(v; n_{i}, T_{i}) dv= f(\omega; n_{i}, T_{i}, \Delta \omega, \omega_{0}) d \omega$ holds with the help of Eq. \eqref{eq:Doppler}.
In our experimental setup, we focus on the spectral line of Ar II whose eigenfrequency and mass are respectively $\omega_{0}=448.37$ THz and $m=6.6905 \times 10^{-26}$ kg, where $n_{i}$, $T_{i}$, and $\Delta \omega$ depend on space (see Methods). We assume that $f(v; n_{i}, T_{i})$ can be observed by the LIF in any other $\omega_{0}$ since $f(v; n_{i}, T_{i})$ is invariant even if $f(\omega; n_{i}, T_{i}, \Delta \omega, \omega_{0})$ is changed by $\omega_{0}$.

\begin{figure*}[h]
\begin{center}
\includegraphics[width=0.5 \textwidth]{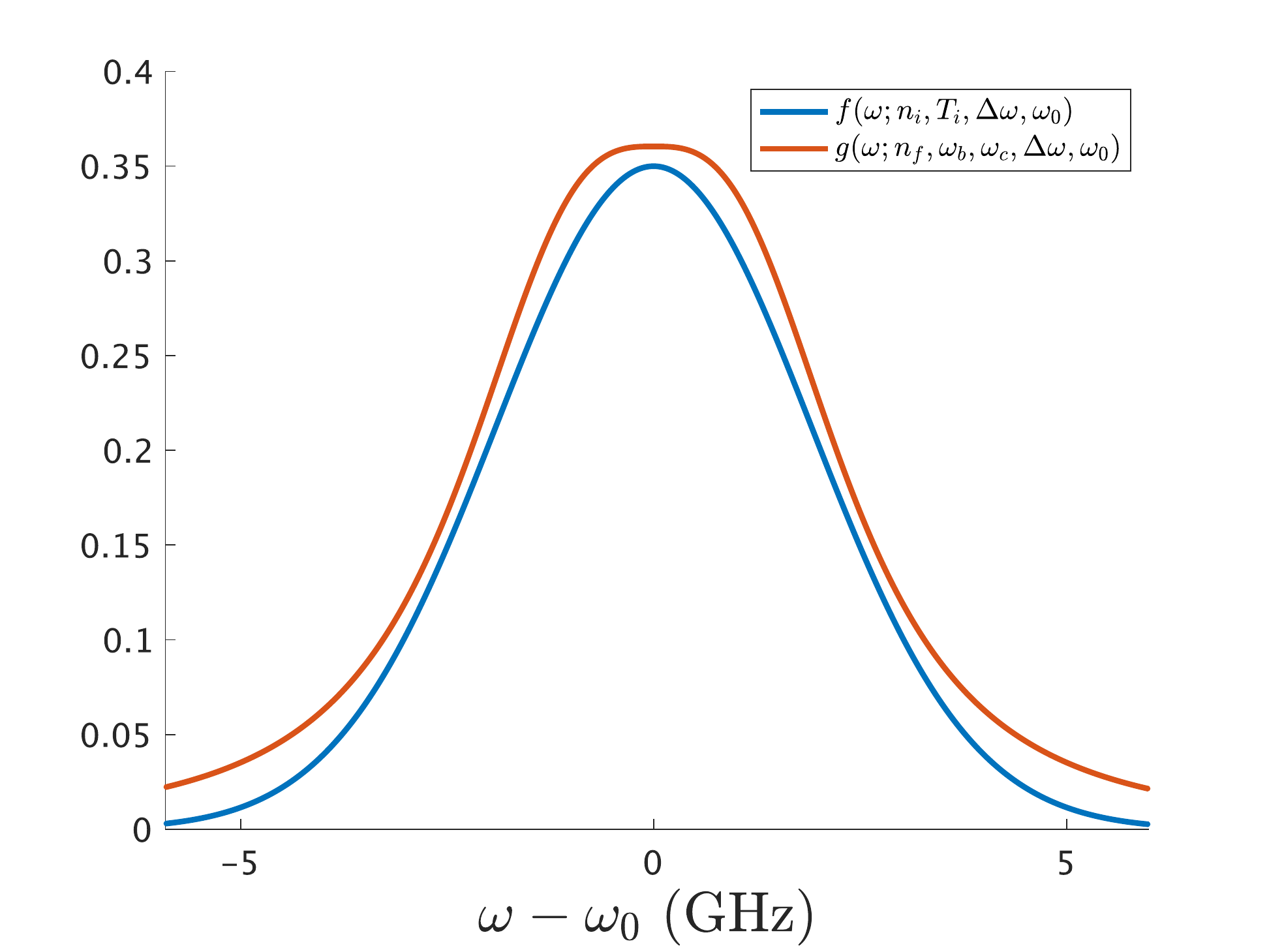}
\end{center}
\caption{Comparison of the Maxwellian and slowing-down line-shapes in the absorption frequency domain. The horizontal axis represents $\omega - \omega_{0}$, where $\omega_{0}=448.37$ GHz.
The Maxwellian line-shape (blue line) is plotted, where $n_{i}=1.7$ (a.u.), $T_{i}=0.68$ eV, and $\Delta \omega = 0$ THz.
The slowing-down line-shape (red) is plotted, where $n_{f}=3.9 \times 10^7$ (a.u.), $\omega_b=-10$ GHz, $\omega_c=-2.4$ GHz, and $\Delta \omega = 0$ GHz. They correspond to best-fit parameters to a typical LIF spectrum of Ar II at the plasma center in our experimental setup.
\label{Fig.1}}
\end{figure*}

If the fast ions at $v=v_b$ are injected into the above system, they will slow down due to the collisions of them with the thermal ions and electrons in the background plasma. Here, we suppose that the density $n_{f}$ of fast ions is much less than the density of thermal ions, and that $v_b$ is much greater than thermal ion velocity but much less than thermal electron velocity. Then, the velocity distribution function of fast ions will reach
\begin{align}
g(v; n_{f}, v_b, v_c):= \frac{3 n_{f}}{4 \log \left( 1+ (v_b/v_c)^3 \right)} \frac{H(v_b-v)}{v^3 + v_c^3}
\label{eq:slowing-down}
\end{align}
at the steady state, where $H(\cdots)$ is the Heaviside step function, and $v_c$ is the crossover velocity \cite{goldston2020introduction, gaffey1976energetic, estrada2006turbulent}. If $v < v_c$, the collisions with the thermal ions is dominant. If $v > v_c$, the collisions with the thermal electrons is dominant. We also obtain the spectral line shape
\begin{align}
g(\omega; n_{f}, \omega_b, \omega_c, \Delta \omega, \omega_{0}):= \frac{3 n_{f} \omega_{0}^2}{4 c^2 \log \left( 1+ |(\omega_b -\Delta \omega - \omega_{0})/(\omega_c -\Delta \omega - \omega_{0})|^3 \right)} \frac{H(|\omega_b -\Delta \omega - \omega_{0}|-|\omega-\Delta \omega - \omega_{0} |)}{|\omega-\Delta \omega - \omega_{0}|^3 + |\omega_c-\Delta \omega - \omega_{0} |^3},
\end{align}
where $\omega_b - \Delta \omega= \omega_{0} (1+v_b/c)$, $\omega_c - \Delta \omega = \omega_{0} (1+v_c/c)$, and $g(v; n_{f}, v_b, v_c) dv = g(\omega; n_{f}, \omega_b, \omega_c, \Delta \omega, \omega_{0}) d \omega$ hold with the help of Eq. \eqref{eq:Doppler}. We should mention that the line shapes of $f(\omega; n_{i}, T_{i}, \Delta \omega, \omega_{0})$ and $g(\omega; n_{f}, \omega_b, \omega_c, \Delta \omega, \omega_{0})$ can be roughly the same for certain parameters and domain (Fig. \ref{Fig.1}). While we expect that there are no fast ions in our experimental setup, it is worth challenging to identify which model represents the observed spectrum more adequately in the statistical sense, $f(\omega; n_{i}, T_{i}, \Delta \omega, \omega_{0})$ or $g(\omega; n_{f}, \omega_b, \omega_c, \Delta \omega, \omega_{0})$.

In magnetized plasmas, the Zeeman effect is not necessarily negligible \cite{boivin2003laser, kelly2016ari, arakawa2019ion}.
The Zeeman effect has the anisotropy dependent on the polarization of laser.  
When the polarization is parallel to the magnetic field, the $\pi$ transition occurs.
When the polarization is perpendicular to the magnetic field, the $\sigma$ transition occurs.
Here, we focus on the $\pi$ transitions, which respectively deform $f(\omega; n_{i}, T_{i}, \Delta \omega, \omega_{0})$ and $g(\omega; n_{f}, \omega_b, \omega_c, \Delta \omega, \omega_{0})$ as 
\begin{align}
\tilde{f}(\omega; n_{i}, T_{i}, \Delta \omega, \omega_{0}, \delta) := \sum_{k=1}^K \pi_k \left( f(\omega; n_{i}, T_{i}, \Delta \omega, \omega_{0} + (2k-1) \delta) + f(\omega; n_{i}, T_{i}, \Delta \omega, \omega_{0} - (2k-1) \delta) \right)
\end{align}
and
\begin{align}
\tilde{g}(\omega; n_{f}, \omega_b, \omega_c, \Delta \omega, \omega_{0}, \delta) := \sum_{k=1}^K \pi_k \left( g(\omega; n_{f}, \omega_b, \omega_c, \Delta \omega, \omega_{0} + (2k-1) \delta) + g(\omega; n_{f}, \omega_b, \omega_c, \Delta \omega, \omega_{0} - (2k-1) \delta) \right),
\end{align}
where $2K$ is the Zeeman component number, $\{\pi_k\}$ are the transition rates, and $2 \delta$ is the Zeeman energy. 
Note that we ignored the dependence of $\Delta \omega$ on $\pm (2k-1) \delta$ since $(2k-1)\delta/\omega_{0} \ll 1$.
In our experimental setup, we consider that $K=3$, $\{\pi_1, \pi_2, \pi_3 \}=\{5/18, 3/18, 1/18 \}$, and $\delta=83.977$ (MHz) for $0.09$ T magnetic field strength \cite{boivin2003laser} (see Methods). 

We should emphasize that $f(\omega; n_{i}, T_{i}, \Delta \omega, \omega_{0})$ and $\tilde{f}(\omega; n_{i}, T_{i}, \Delta \omega, \omega_{0}, \delta)$ are different in the internal degrees of freedom, $\omega$ or $\omega_{0} \pm (2k-1) \delta$, while they reflect the same statistical behavior of ionic motion subject to $f(v; n_{i}, T_{i})$. The case of $g(\omega; n_{f}, \omega_b, \omega_c, \Delta \omega, \omega_{0})$ and $\tilde{g}(\omega; n_{f}, \omega_b, \omega_c, \Delta \omega, \omega_{0}, \delta)$ is too. If an observed spectrum is deformed, it is not always trivial to identify which factor is the origin of such a deformation, the ionic motion or the internal degrees of freedom. This ambiguity makes it more complicated to identify what factor dominates ionic motion from the observed spectrum.
Under these difficulties, we statistically evaluate which form of the spectral intensity function $I(\omega; w)$, as listed in Table \ref{tab:function_{f}orm}, is adequate for the observed LIF spectrum, where $I_0$ is the background intensity.
This evaluation links to identify the corresponding form of the velocity distribution function.

While many other mechanisms contribute to the spectral line broadening in LIF, such as natural broadening, power broadening, Stark broadening, and instrumental broadening, we ignore them for simplicity. This is because our experimental setup is almost the same as the setup of Ref. \cite{boivin2003laser}, in which the Doppler and Zeeman effects are dominant in order. There are also many other models of ion velocity distribution function in magnetized plasma, as reviewed in Ref. \cite{moseev2019bi}. Depending on the setup, they can also be candidates, not only $f(v; n_{i}, T_{i})$ and $g(v; n_{f}, v_b, v_c)$, to be evaluated in terms of the statistical adequateness for the observed spectrum.
\\

\begin{figure}[h]
\begin{center}
    \begin{tabular}{c}
         \includegraphics[width=0.5 \textwidth]{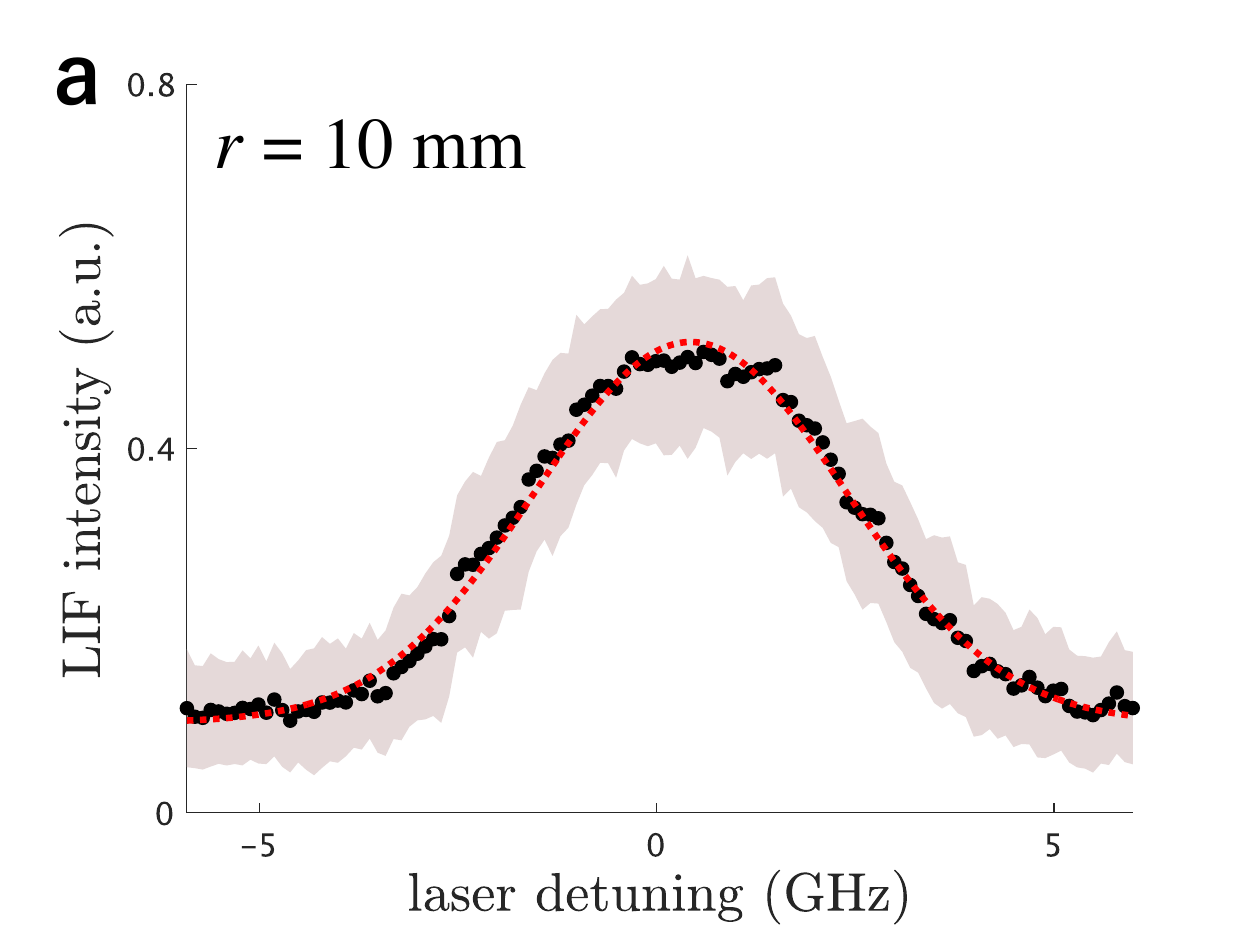}  \\
         \includegraphics[width=0.5 \textwidth]{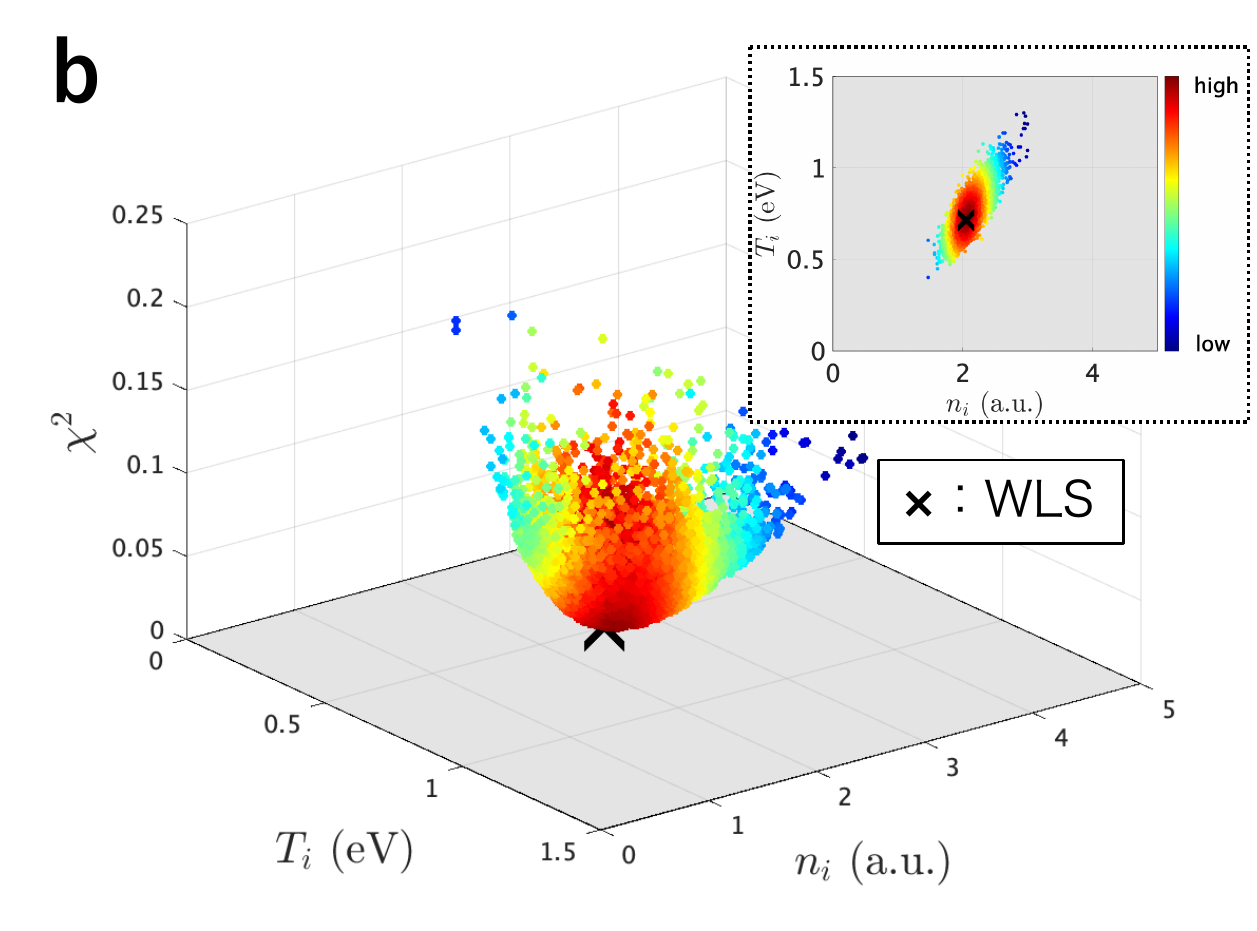}
    \end{tabular}
\end{center}
\caption{Data, model, and its parameters. (a) Curve fitting (red dashed line) to a typical LIF spectrum (black dots), assuming a Maxwellian distribution (function II). This analysis is based on WLS by considering the error bars (shade) at all data points. (b) Scatter plot of a "statistical ensemble" (colour dots) in the Bayesian inference based on the assumption above, compared with the WLS solution (black cross). The colour of each dot represents the posterior probability density, which the inset shows as a projection in the direction of $\chi^2$.\label{Fig.2}}
\end{figure}

\noindent
\textbf{Basics of Bayesian inference.}
We explain the basic concept of Bayesian inference, comparing it with the conventional LIF analysis.
A common approach in measuring the ion temperature and the ion flow velocity in steady-state plasmas is to find the best reproduction of a LIF spectrum by the spectral intensity function $I(\omega; w) = f(\omega; n_{i}, T_{i}, \Delta \omega) + I_0$ (function II in Table \ref{tab:function_{f}orm}) with the parameter set $w=\{n_{i}, T_{i}, \Delta \omega, I_0\}$. A typical LIF spectrum of Ar II at the radial position $r=10$ mm in a cylindrical helicon plasma (see Methods) and its reproduction are shown in Fig. \ref{Fig.2}(a).
A best-fit parameter set $w$ for a LIF spectrum is obtained by the weighted least squares (WLS), namely minimization of the chi-squared function
\begin{align}
\chi^2(w; D^n) := \frac{1}{n} \sum_{j=1}^{n} \frac{(y_j - I(\omega_j; w))^2}{\sigma_j^2}
\end{align}
of $w$ given the data set $D^n := \{y_j, \omega_j, \sigma_j\}_{j=1}^n$, where $y_j$ and $\sigma_j$ are respectively the average and the standard deviation of LIF intensity over several discharges at $\omega_j$.

Now we extend this common approach to a Bayesian inference. Suppose that $y_1, y_2, \cdots, y_n$ are independent samples taken from each conditional probability distribution  
\begin{align}
p(y \mid \omega_j, \sigma_j, w) := \frac{1}{\sqrt{2 \pi \sigma_j^2}} \exp \left( - \frac{(y - I(\omega_j; w) )^2}{2 \sigma_j^2}\right).
\label{eq:noise}
\end{align}
In other words, if a random variable $Y_j$ is subject to $p(y \mid \omega_j, \sigma_j, w)$, then it satisfies the relation $Y_j = I(\omega_j; w) + \xi_j$, where $\xi_j$ is a noise subject to the Gaussian distribution $\mathcal{N}(0,\sigma_j^2)$.
It is considered that $Y_1, Y_2, \cdots, Y_n$ are a stochastic process since their realizations $y_1, y_2, \cdots, y_n$ are originally a time-series (see Methods). Namely, $\xi_1, \xi_2, \cdots, \xi_n$ are also a stochastic process.
We expect that Eq. \eqref{eq:noise} is valid if $Y_j$ is stationary, i.e., $\xi_j$ is white, since the conditional independence of $Y_1, Y_2, \cdots, Y_n$, which Eq. \eqref{eq:noise} requires, is supported. In our experimental setup, we set the step size $\omega_{j+1} - \omega_{j}=0.1$ GHz for any $j$ to sufficiently satisfy the requirement of Eq. \eqref{eq:noise} by checking the correlation length of raw data in the time domain.

In the Bayesian approach, $w$ is also regarded as a random variable subject to the conditional probability distribution
\begin{align}
p(w \mid D^n) &= \frac{p(w)}{p(\{y_j\}_{j=1}^n \mid \{\omega_j, \sigma_j\}_{j=1}^n)} \prod_{j=1}^n p(y_j \mid \omega_j, \sigma_j, w) \notag \\
&\propto \exp \left( -\frac{n}{2} \chi^2(w; D^n) \right),
\label{eq:posterior}
\end{align}
where $p(w)$ and $p(\{y_j\}_{j=1}^n \mid \{\omega_j, \sigma_j\}_{j=1}^n)$ are respectively the uniform distribution as our working hypothesis and the normalizing constant.  
Since Eq. \eqref{eq:posterior} is derived from Bayes' formula, $p(w)$ and $p(w \mid D^n)$ are respectively called the prior and posterior probability density functions.

By following the mathematical correspondence between Bayesian inference and statistical mechanics \cite{jaynes2003probability, balasubramanian1997statistical, zdeborova2016statistical}, the form of $p(w \mid D^n)$ is a "Boltzmann distribution" whose "inverse temperature" and "energy" for a "state" $w$ are respectively $n$ and $\chi^2(w; D^n)$ as an analogy. Whereas WLS is to obtain a "ground state" as a best-fit parameter set, Bayesian inference is to obtain a whole "statistical ensemble" subject to $p(w \mid D^n)$ (or to obtain $p(w \mid D^n)$ itself).
A "statistical ensemble", namely numerous realizations of $w$, subject to $p(w \mid D^n)$ is shown in Fig. \ref{Fig.2}(b).
One can see the posterior probability density $p(n_{i}, T_{i} \mid D^n)$, represented by the colour gradation in the inset, tends to be higher as $\chi^2(w; D^n)$ is smaller, where WLS solution is located around the highest density.
This tendency is supported by Eq. \eqref{eq:posterior}; WLS solution corresponds to $w$ that maximize $p(w \mid D^n)$.
Note that this explanation is not strict but intuitive since the two values of $w$ that respectively maximize $p(w \mid D^n)$ and $p(n_{i}, T_{i} \mid D^n)$ are almost same but strictly different \cite{footnote1}. 
In the Bayesian approach, the "ensemble" average and standard deviation are respectively taken as one of the estimators and its error bar, referred to as the posterior mean and standard deviation, since the $w$'s "fluctuation" means the uncertainty in parameter estimation: e.g. $n_{i} = 2.0 \pm 0.2$ (a.u.) and $T_{i} = 0.74 \pm 0.11$ (eV) for Fig. \ref{Fig.2}(b).
\\

\begin{figure}[h]
\begin{center}
\includegraphics[width=0.5 \textwidth]{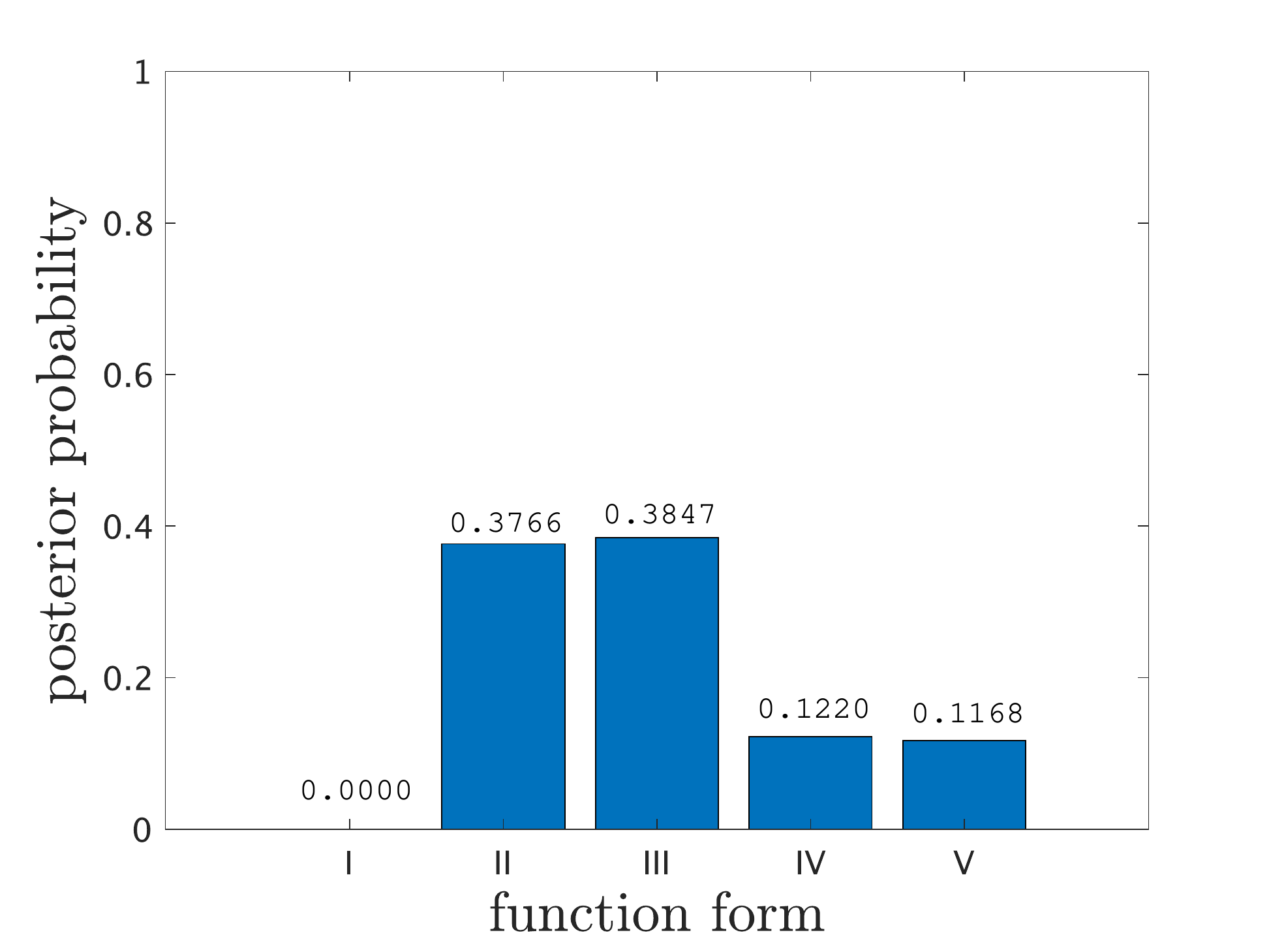}
\end{center}
\caption{Posterior probability against form of ion velocity distribution function. The assumed forms are no ions (function I), a Maxwellian (function II), a Maxwellian with Zeeman splitting (function III), a slowing-down (function IV), and a slowing-down with Zeeman splitting (function V). \label{Fig.3}}
\end{figure}

\noindent
\textbf{Bayesian model selection.}
We address the question of what form of ion velocity distribution function is adequate to describe a observed spectrum in a statistical sense. In the Bayesian approach, a relative goodness of each $I(\omega; w)$'s form, labeled by $M_l$, for $D^n$ is quantified by the conditional probability 
\begin{align}
p(M_l \mid D^n) &= \frac{p(\{y_j\}_{j=1}^n \mid \{\omega_j, \sigma_j\}_{j=1}^n, M_l)}{\sum_l p(\{y_j\}_{j=1}^n \mid \{\omega_j, \sigma_j\}_{j=1}^n, M_l)},
\label{eq:posterior_model}
\end{align}
where 
\begin{align}
p(\{y_j\}_{j=1}^n \mid \{\omega_j, \sigma_j\}_{j=1}^n, M_l) &= \int p(w \mid M_l) \prod_{j=1}^n p(y_j \mid \omega_j, \sigma_j, w, M_l) dw
\end{align}
is the normalizing constant in Eq. \eqref{eq:posterior}, explicitly representing the dependence on $M_l$, such as $p(w \mid M_l)$ for $p(w)$. Note that Eq. \eqref{eq:posterior_model} is derived from Bayes' formula such that $p(M)$ for $M \in \{M_l\}$ is the uniform distribution.
We calculate $p(M_l \mid D^n)$ for the five different function forms listed in Table \ref{tab:function_{f}orm}, where $D^n$ is the same spectrum shown in Fig. \ref{Fig.2}(a).
This result links to verify that the ion velocity distribution corresponding to the observed spectrum is a Maxwellian rather than a slowing-down.
One can see from Fig. \ref{Fig.3} that function III has the highest probability among the five forms. This probability distribution supports that the function expected to be physically valid is also statistically valid in this case. One might concern that function II ($37.66$\%) is significantly competing with function III ($38.47$\%), where their difference is the presence of the Zeeman effect. This competition implies that the fine structure is not easy to be distinguished from the gross structure, but the observed LIF spectrum barely has the quality to do so.
The posterior probability of function I significantly lower than the two above supports that there are Ar II ions at the radial position $r=10$ mm.

\begin{figure*}[h]
\includegraphics[width=1 \textwidth]{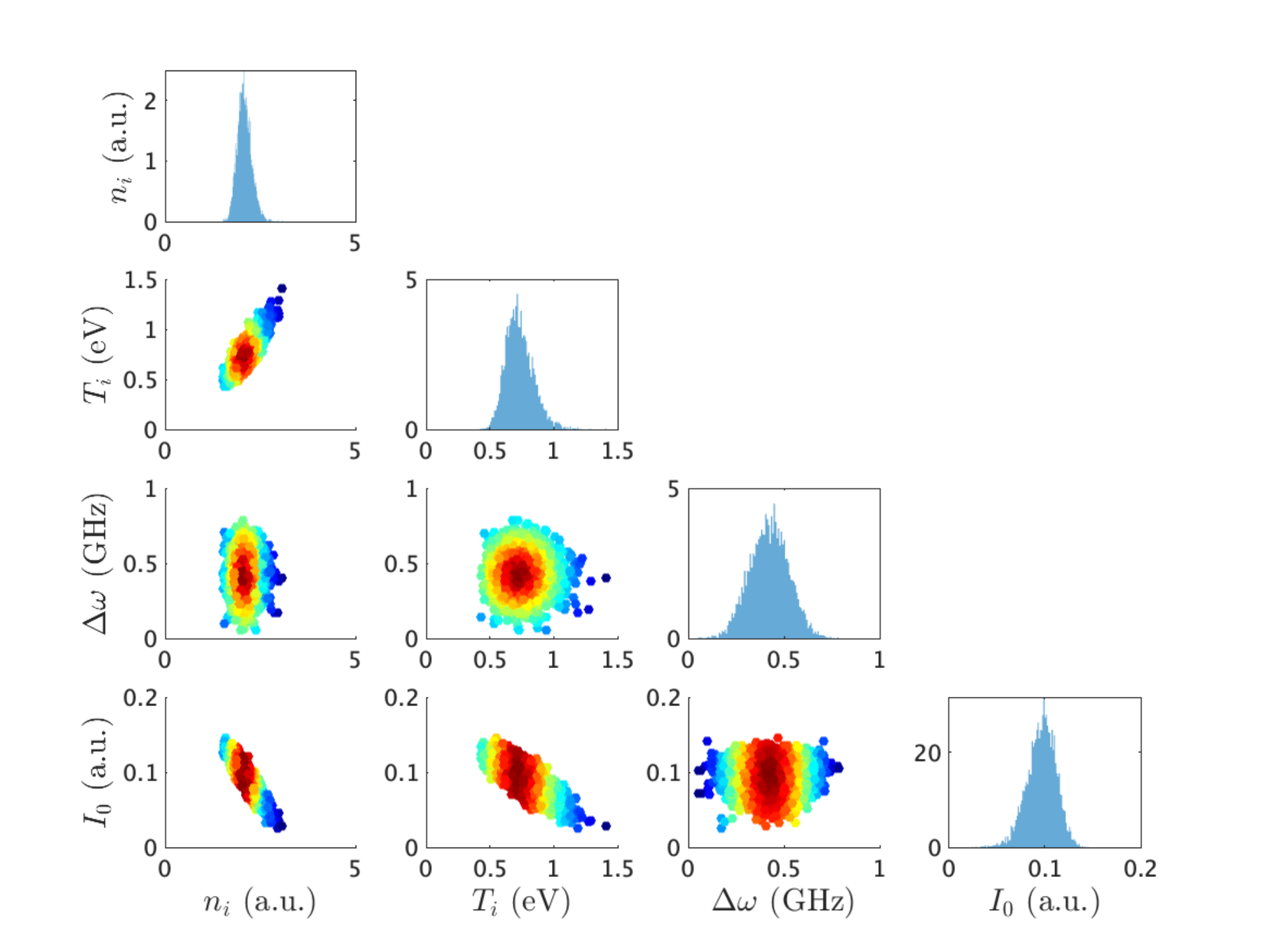}
\caption{Corner plot of posterior probability density under the assumption of function III. The panels on the diagonal show the histogram as the posterior probability density for each parameter, obtained by marginalizing over the other parameters. The off-diagonal panels show the scatter plots as the posterior probability distributions for each pair of parameters. The colour of each dot represents posterior probability density on each panel. \label{Fig.4}}
\end{figure*}

One can further confirm the uncertainty in parameter estimation. Each panel on the diagonal of Fig. \ref{Fig.4} signifies that the marginal posterior density function of each parameter converges to Gaussian distribution. Such a convergence indicates that the amount and quality of the observed LIF spectrum are enough.
Each panel on the off-diagonal of Fig. \ref{Fig.4} signifies there is a correlation between each pair of parameters. Such a correlation comes from the form of $I(\omega; w)$, which characterizes $p(w \mid D^n)$.
\\

\begin{figure*}[h]
\begin{center}
    \begin{tabular}{ccc}
         \includegraphics[width=0.32 \textwidth]{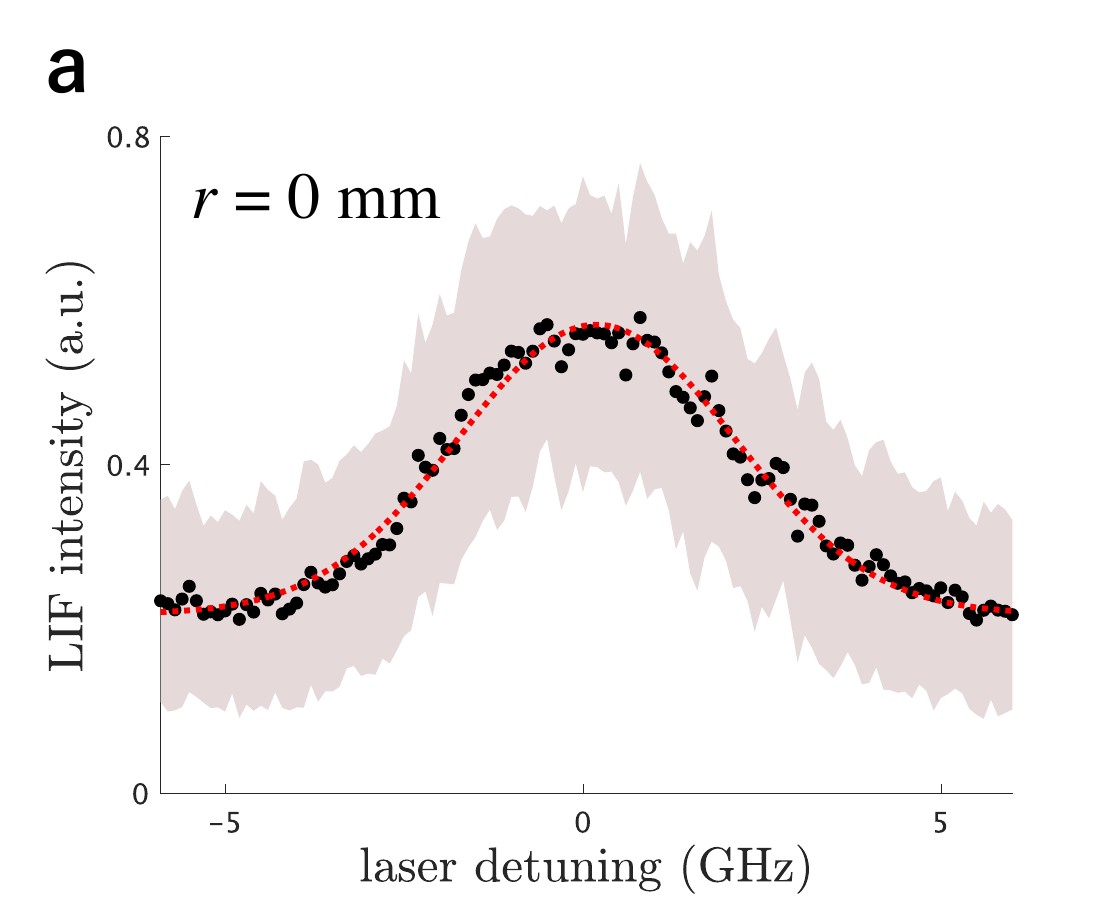} & 
         \includegraphics[width=0.32 \textwidth]{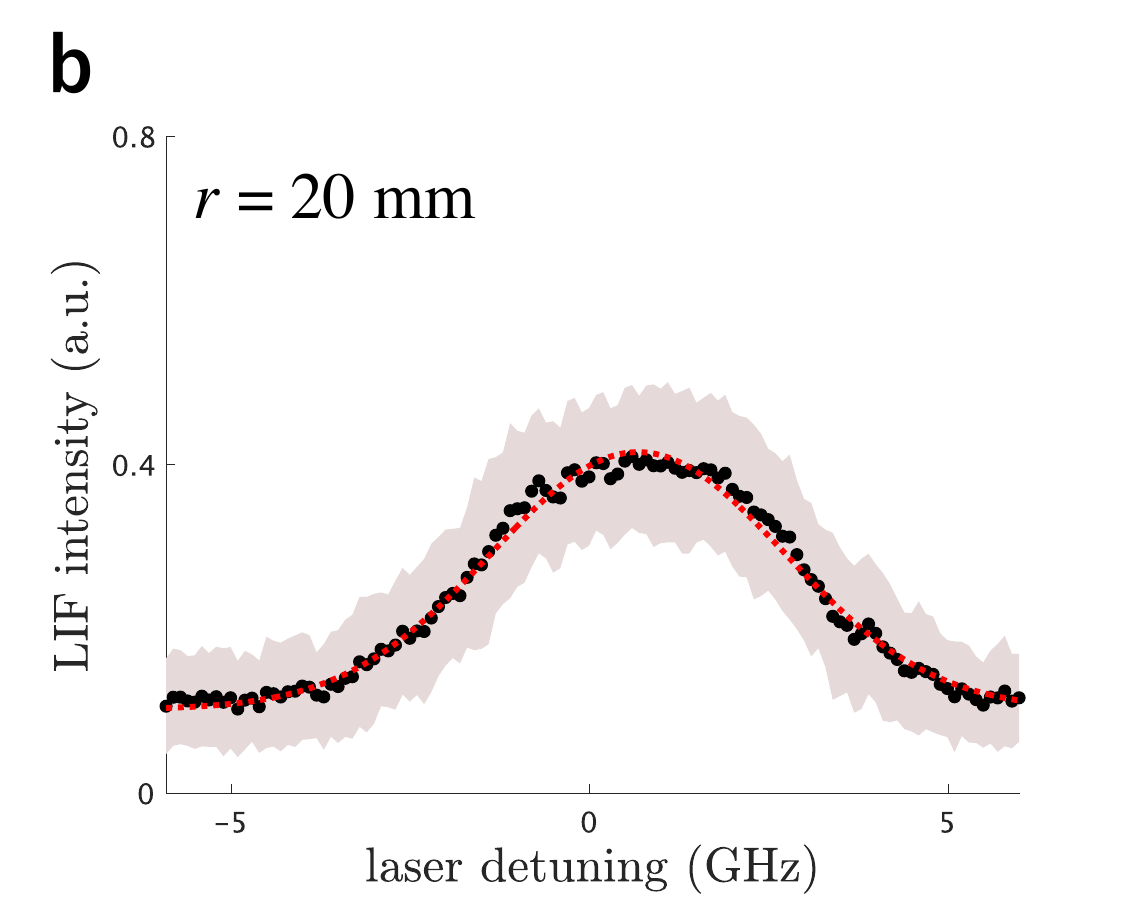} &
         \includegraphics[width=0.32 \textwidth]{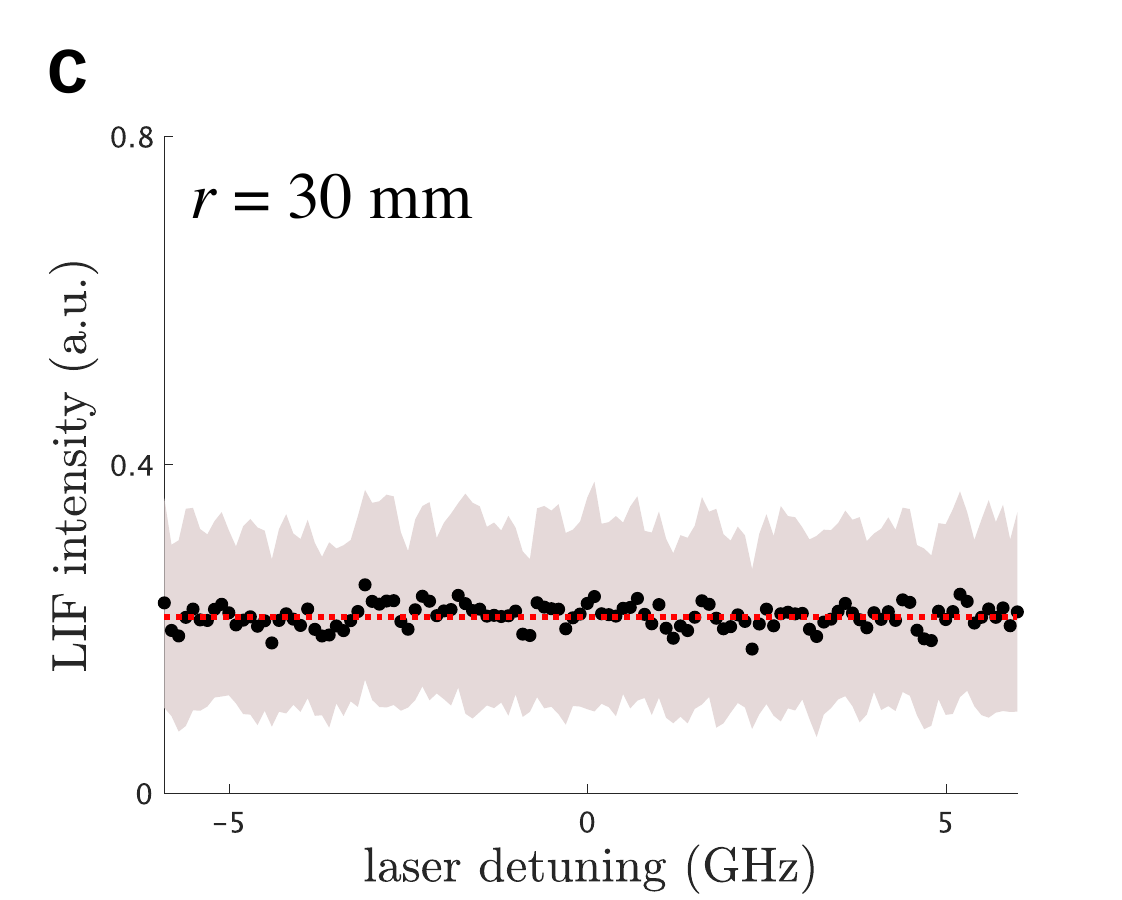} \\
         \includegraphics[width=0.32 \textwidth]{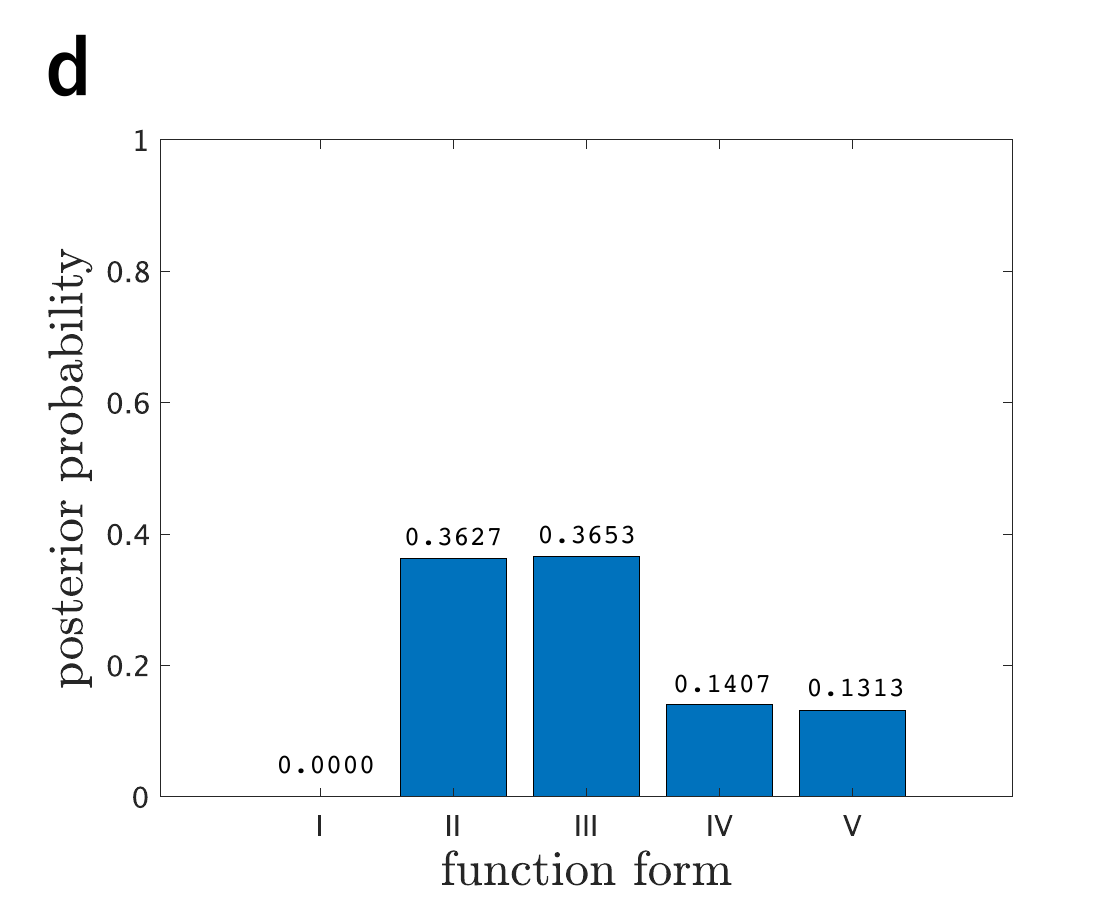} & 
         \includegraphics[width=0.32 \textwidth]{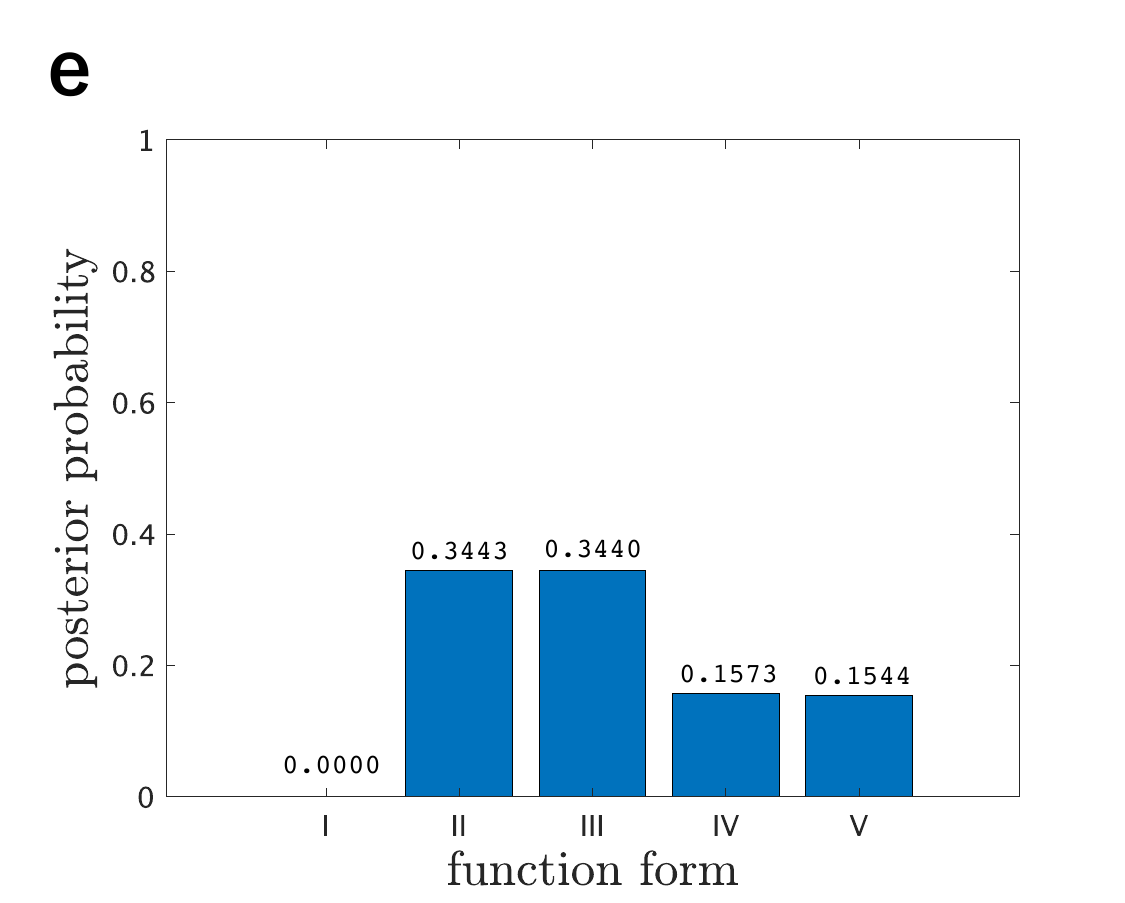} &
         \includegraphics[width=0.32 \textwidth]{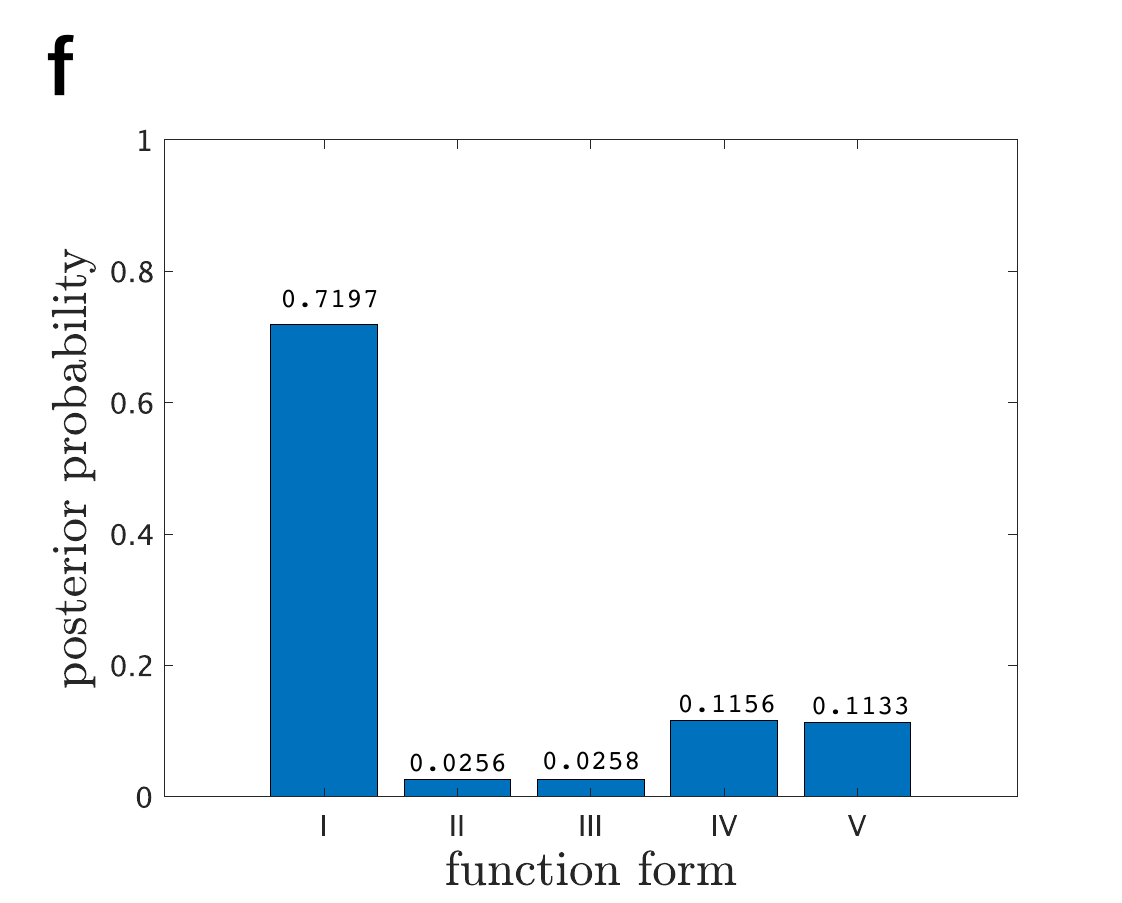}

    \end{tabular}
\end{center}
\caption{Bayesian model selection for LIF spectrum at each radius. (a-c) LIF spectra (black dots) and their reproduction (red dashed line) at $r=0$ mm, $20$ mm, and $r=30$ mm. The reproduction curves are best-fit solutions assuming function III for $r=0$ mm and $20$ mm, and function I for $r=30$ mm. (d-f) Posterior probability against the form of ion velocity distribution function under LIF spectrum at each radius. \label{Fig.5}}
\end{figure*}

\noindent
\textbf{Evaluation of phase-space distribution.}
We extend our approach to evaluate the spatial inhomogeneity in a cylindrical helicon plasma. The same analysis as above is additionally performed for each $D^n$ as observed LIF spectrum at each radial position $0$ mm, $20$ mm, and $30$ mm (Fig. \ref{Fig.5}).
One can see from Fig. \ref{Fig.5}(d-f) that the valid form, whose $p(M_l \mid D^n)$ is highest for each radial position, is function III for $0$ mm, function II for $20$ mm, and function I for $30$ mm. The first correspond to a straightforward account of physics concerned with the experimental setup; there is the Zeeman effect induced by the external homogeneous magnetic field (Fig. \ref{Fig.3}). The second indicates that the Zeeman effect is sometimes negligible, as well as other types of line broadening. One might concern a non-straightforward result that function I is valid for $30$ mm. This result is a lesson that WLS solution is just an unreliable value for ion temperature and flow velocity if one straightforwardly assume function II or III for invalid situations.

\begin{figure*}[h]
\begin{center}
    \begin{tabular}{ccc}
         \includegraphics[width=0.32 \textwidth]{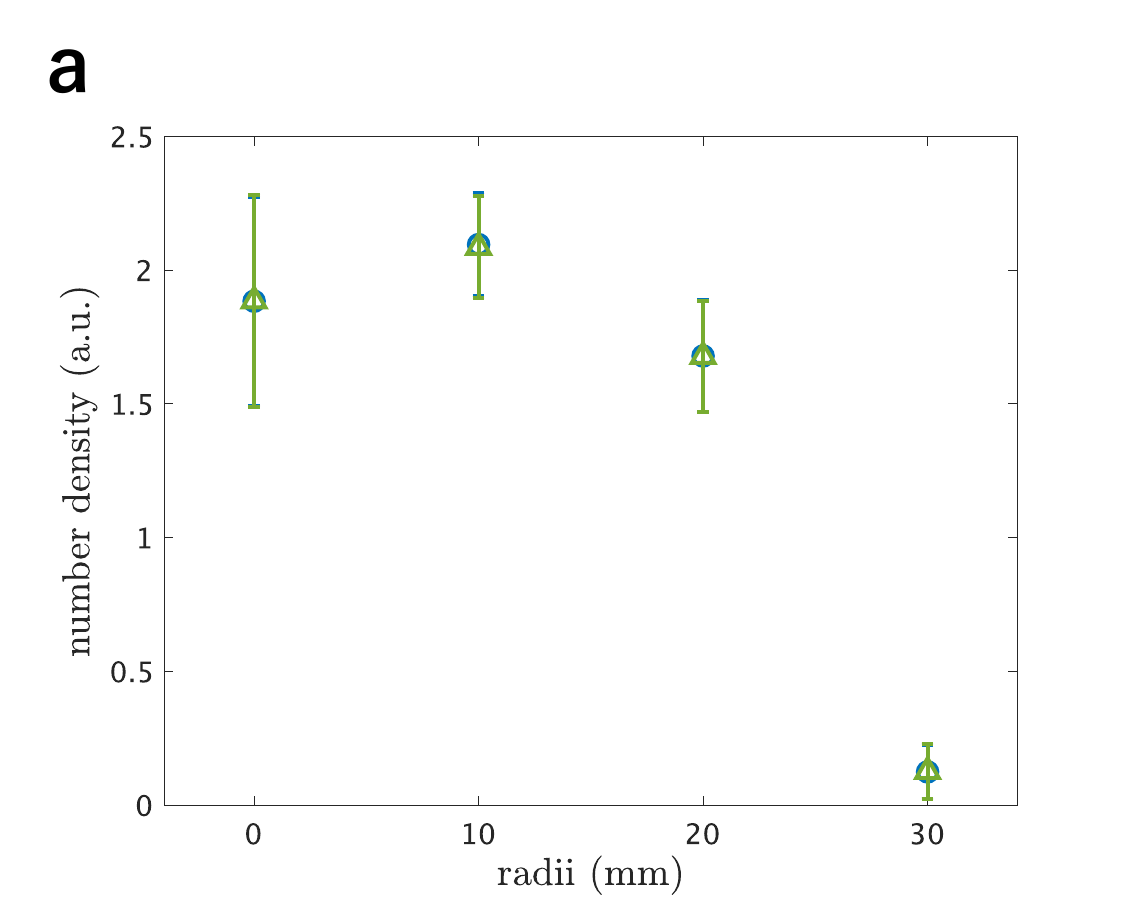} & 
         \includegraphics[width=0.32 \textwidth]{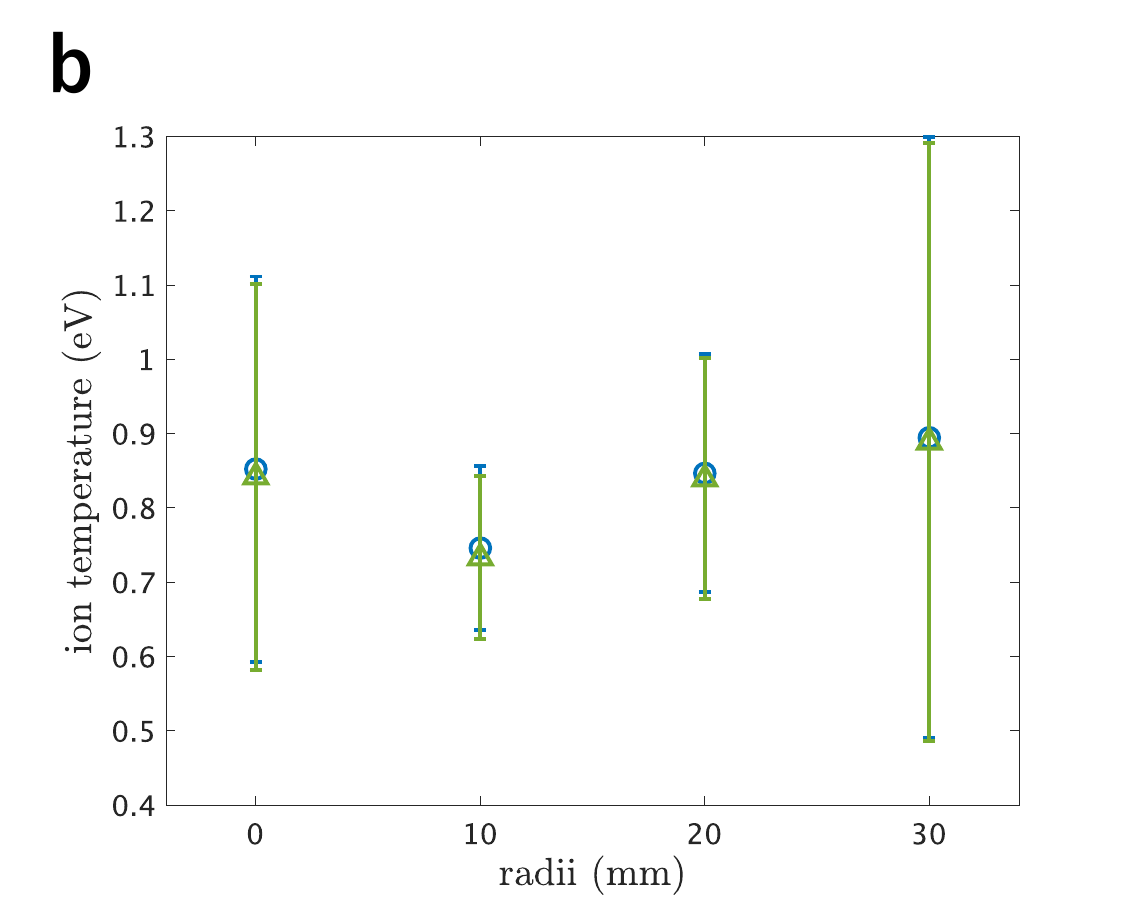} &
         \includegraphics[width=0.32 \textwidth]{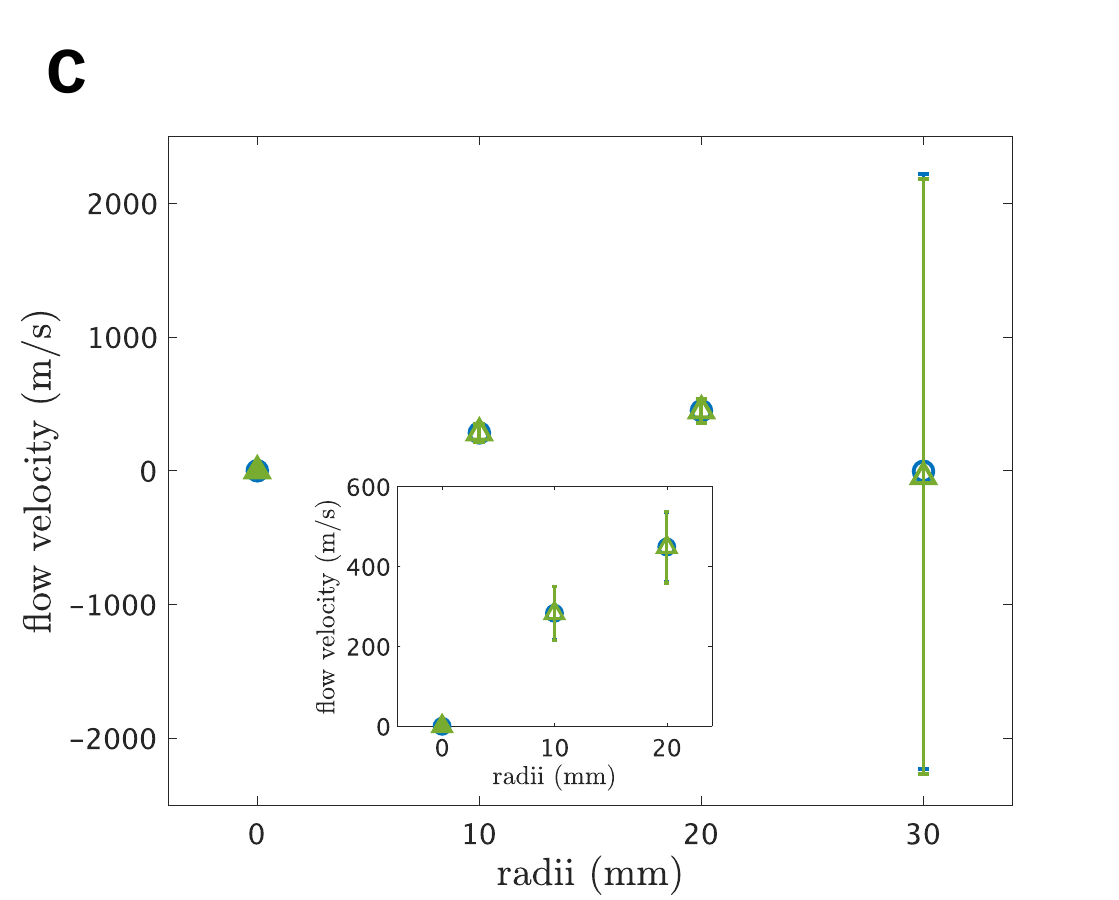}
    \end{tabular}
\end{center}
\caption{Spatial inhomogeneity in cylindrical helicon plasma. The error bars represent posterior standard deviation. (a) Plots of ion density against radii, respectively obtained by assuming functions II (blue circle) and III (green triangle). (b) Plots of ion temperature against radii. (c) Plots of ion flow velocity against radii. Inset shows the magnified view. \label{Fig.6}}
\end{figure*}

We plot in Fig. \ref{Fig.6} the radial profiles of (a) ion density $n_{i}$, (b) ion temperature $T_{i}$, and (c) flow velocity $c\Delta \omega/\omega_{0}$, where $\Delta \omega =0$ is fixed for $r=0$ mm by considering the symmetry.
Two different profiles are respectively obtained under assumptions of functions II and  III for comparison. They are fairly consistent, including their error bars. This consistency is connected with the fact that the posterior probabilities of functions II and III are almost the same; the Zeeman effect is sometimes negligible. One might concern significantly large errors at $30$ mm, shown in Fig. \ref{Fig.6}(c) and Fig. \ref{Fig.6}(c), corresponding to $n_{i} \simeq 0$ (Fig. \ref{Fig.6}(a)). Such large uncertainties reflects that $\Delta \omega$ and $T_{i}$ are arbitrary in the case $n_{i} \simeq 0$, namely $I(\omega;w) \simeq I_0$. This result is also supported by the fact that function I is valid form of the ion velocity distribution (Fig. \ref{Fig.5}(f)).

\section*{Discussion}
The present study sheds light on the ambiguity in identifying the velocity distribution function out of thermodynamic equilibrium states from LIF spectra. We demonstrate that our approach can resolve the ambiguity. The success of our approach relies on reliable LIF measurements. Our approach is also applicable to other spectroscopic techniques, such as absorption spectroscopy and two photons absorption LIF \cite{demtroder1973laser}.

Our approach may not work as expected if the signal-to-noise ratio (SNR) is not sufficiently high. The uncertainty in parameters and function forms is connected with the SNR in the Bayesian inference. It is pointed out that there are thresholds of the SNR permissible for parameter estimation \cite{nagata2019bayesian} and model selection \cite{tokuda2017simultaneous} in similar contexts. 
The systematic error from experimental artefacts, e.g., the thermal drift of laser and the reproducibility of plasma production, may also bring unexpected results since we suppose that there are only random errors in our formulation (Eq. \eqref{eq:noise}).
The plasma instabilities and fluctuation may be sources of error depending on the setup. In order to obtain more reliable results, it is essential to minimize these systematic errors and disturbances.

\section*{Methods}
\textbf{PANTA experiments.} The experiment was carried out in a linear magnetized plasma device, PANTA\cite{inagaki2016concept}.
The vacuum vessel has the length of 4050 mm and the diameter of 450 mm.
0.09 T of homogeneous magnetic field  is controlled by 17 pairs of Helmholtz coils.
Argon plasma with 50 mm radius is produced by helicon wave with 3\,kW at 7\,MHz radio frequency, and neutral gas pressure is fixed at 0.1 Pa, where the streamer structure are observed \cite{yamada2008anatomy}.
The discharge duration is 2.0 s.
Typical central electron density and temperature are $\sim 1 \times 10^{19} \mathrm{m^{-3}}$ and 3\,eV, respectively \cite{tomita2017measurement}.
\\

\noindent
\textbf{LIF measurements.} Here, we briefly review the LIF measurement system in PANTA.
A tunable diode laser tuned at around 668.61 nm is injected perpendicular to the magnetic fields and utilized for exciting the $3d^4 F_{7/2}$ level to the $4p^4 D_{5/2}$ level in Ar $\mathrm{I\hspace{-.1em}I}$ in the experiments\cite{severn1998argon}.
The laser wavenumber is swept around 668.61$\pm$0.01 nm by triangle waveform with 0.1 Hz of modulation frequency.
The effect of laser thermal drift has been calibrated using a Fabry-P\'{e}rot interferometer and an iodine cell.
The LIF signal is collected by a photo multiplier tube and amplify and averaging by lock-in amplifier.
Since the error of the LIF signal is large, we performed 180 discharges at each radial position in the experiments
The LIF signals at 0.2 s after the start of discharge, when the turbulence becomes quasi-steady state, is used to evaluate the ion velocity distribution function.
For more details of the LIF system, see Ref. \cite{arakawa2019ion}.
\\

\noindent
\textbf{Monte Carlo simulations.} Throughout this study, the support of $p(w)$, namely the domain of $w$, is set as follows: $n_{i} \in [0, 5]$ (a.u.), $T_{i} \in [0, 1.5]$ (eV), $n_{f} \in [0, 5 \times 10^7]$ (a.u.), $\omega_b \in [-12 + \omega_{0}, 12+ \omega_{0}]$ (GHz), $\omega_c \in [-6 + \omega_{0}, 6+ \omega_{0}]$ (GHz), $\Delta \omega \in [-6, 6]$ (GHz), and $I_0 \in [0, 1]$ (a.u.). To sample $w$ from $p(w \mid D^n)$, we performed Monte Carlo (MC) simulations by using the parallel tempering based on the Metropolis criterion \cite{geyer1991markov, hukushima1996exchange}, where the total MC sweeps were 10,000 after the burn-in.
To calculate $p(\{y_j\}_{j=1}^n \mid \{\omega_j, \sigma_j\}_{j=1}^n)$, we utilized the bridge sampling \cite{meng1996simulating, gelman1998simulating}.

\bibliography{main.bbl}

\begin{thebibliography}{10}
\expandafter\ifx\csname url\endcsname\relax
  \def\url#1{\texttt{#1}}\fi
\expandafter\ifx\csname urlprefix\endcsname\relax\def\urlprefix{URL }\fi
\providecommand{\bibinfo}[2]{#2}
\providecommand{\eprint}[2][]{\url{#2}}

\bibitem{goldston2020introduction}
\bibinfo{author}{Goldston, R.~J.}
\newblock \emph{\bibinfo{title}{Introduction to plasma physics}}
  (\bibinfo{publisher}{CRC Press}, \bibinfo{year}{2020}).

\bibitem{moseev2019bi}
\bibinfo{author}{Moseev, D.} \& \bibinfo{author}{Salewski, M.}
\newblock \bibinfo{title}{Bi-maxwellian, slowing-down, and ring velocity
  distributions of fast ions in magnetized plasmas}.
\newblock \emph{\bibinfo{journal}{Physics of Plasmas}}
  \textbf{\bibinfo{volume}{26}}, \bibinfo{pages}{020901}
  (\bibinfo{year}{2019}).

\bibitem{spitzer2006physics}
\bibinfo{author}{Spitzer, L.}
\newblock \emph{\bibinfo{title}{Physics of fully ionized gases}}
  (\bibinfo{publisher}{Courier Corporation}, \bibinfo{year}{2006}).

\bibitem{stix1972heating}
\bibinfo{author}{Stix, T.~H.}
\newblock \bibinfo{title}{Heating of toroidal plasmas by neutral injection}.
\newblock \emph{\bibinfo{journal}{Plasma Physics}}
  \textbf{\bibinfo{volume}{14}}, \bibinfo{pages}{367} (\bibinfo{year}{1972}).

\bibitem{severn1998argon}
\bibinfo{author}{Severn, G.}, \bibinfo{author}{Edrich, D.} \&
  \bibinfo{author}{McWilliams, R.}
\newblock \bibinfo{title}{Argon ion laser-induced fluorescence with diode
  lasers}.
\newblock \emph{\bibinfo{journal}{Review of scientific instruments}}
  \textbf{\bibinfo{volume}{69}}, \bibinfo{pages}{10--15}
  (\bibinfo{year}{1998}).

\bibitem{boivin2003laser}
\bibinfo{author}{Boivin, R.} \& \bibinfo{author}{Scime, E.}
\newblock \bibinfo{title}{Laser induced fluorescence in {A}r and {H}e plasmas
  with a tunable diode laser}.
\newblock \emph{\bibinfo{journal}{Review of scientific instruments}}
  \textbf{\bibinfo{volume}{74}}, \bibinfo{pages}{4352--4360}
  (\bibinfo{year}{2003}).

\bibitem{claire2006laser}
\bibinfo{author}{Claire, N.}, \bibinfo{author}{Bachet, G.},
  \bibinfo{author}{Stroth, U.} \& \bibinfo{author}{Doveil, F.}
\newblock \bibinfo{title}{Laser-induced-fluorescence observation of ion
  velocity distribution functions in a plasma sheath}.
\newblock \emph{\bibinfo{journal}{Physics of plasmas}}
  \textbf{\bibinfo{volume}{13}}, \bibinfo{pages}{062103}
  (\bibinfo{year}{2006}).

\bibitem{kelly2016ari}
\bibinfo{author}{Kelly, R.}, \bibinfo{author}{Meaney, K.},
  \bibinfo{author}{Gilmore, M.}, \bibinfo{author}{Desjardins, T.} \&
  \bibinfo{author}{Zhang, Y.}
\newblock \bibinfo{title}{Ari/arii laser induced fluorescence system for
  measurement of neutral and ion dynamics in a large scale helicon plasma}.
\newblock \emph{\bibinfo{journal}{Review of Scientific Instruments}}
  \textbf{\bibinfo{volume}{87}}, \bibinfo{pages}{11E560}
  (\bibinfo{year}{2016}).

\bibitem{arakawa2019ion}
\bibinfo{author}{Arakawa, H.} \emph{et~al.}
\newblock \bibinfo{title}{Ion temperature measurement by laser-induced
  fluorescence spectroscopy in {PANTA}}.
\newblock \emph{\bibinfo{journal}{IEEJ Transactions on Electrical and
  Electronic Engineering}} \textbf{\bibinfo{volume}{14}},
  \bibinfo{pages}{1450--1454} (\bibinfo{year}{2019}).

\bibitem{efron1973stein}
\bibinfo{author}{Efron, B.} \& \bibinfo{author}{Morris, C.}
\newblock \bibinfo{title}{Stein's estimation rule and its competitors―an
  empirical bayes approach}.
\newblock \emph{\bibinfo{journal}{Journal of the American Statistical
  Association}} \textbf{\bibinfo{volume}{68}}, \bibinfo{pages}{117--130}
  (\bibinfo{year}{1973}).

\bibitem{akaike1980likelihood}
\bibinfo{author}{Akaike, H.}
\newblock \bibinfo{title}{Likelihood and the bayes procedure}.
\newblock \emph{\bibinfo{journal}{Trabajos de Estadistica Y de Investigacion
  Operativa}} \textbf{\bibinfo{volume}{31}}, \bibinfo{pages}{143--166}
  (\bibinfo{year}{1980}).

\bibitem{mackay1992bayesian}
\bibinfo{author}{MacKay, D.~J.}
\newblock \bibinfo{title}{Bayesian interpolation}.
\newblock \emph{\bibinfo{journal}{Neural computation}}
  \textbf{\bibinfo{volume}{4}}, \bibinfo{pages}{415--447}
  (\bibinfo{year}{1992}).

\bibitem{bishop2006pattern}
\bibinfo{author}{Bishop, C.~M.}
\newblock \emph{\bibinfo{title}{Pattern recognition and machine learning}}
  (\bibinfo{publisher}{springer}, \bibinfo{year}{2006}).

\bibitem{demtroder1973laser}
\bibinfo{author}{Demtr{\"o}der, W.}
\newblock \emph{\bibinfo{title}{Laser spectroscopy}}, vol.~\bibinfo{volume}{5}
  (\bibinfo{publisher}{Springer}, \bibinfo{year}{1973}).

\bibitem{gaffey1976energetic}
\bibinfo{author}{Gaffey, J.~D.}
\newblock \bibinfo{title}{Energetic ion distribution resulting from neutral
  beam injection in tokamaks}.
\newblock \emph{\bibinfo{journal}{Journal of Plasma Physics}}
  \textbf{\bibinfo{volume}{16}}, \bibinfo{pages}{149--169}
  (\bibinfo{year}{1976}).

\bibitem{estrada2006turbulent}
\bibinfo{author}{Estrada-Mila, C.}, \bibinfo{author}{Candy, J.} \&
  \bibinfo{author}{Waltz, R.}
\newblock \bibinfo{title}{Turbulent transport of alpha particles in reactor
  plasmas}.
\newblock \emph{\bibinfo{journal}{Physics of Plasmas}}
  \textbf{\bibinfo{volume}{13}}, \bibinfo{pages}{112303}
  (\bibinfo{year}{2006}).

\bibitem{jaynes2003probability}
\bibinfo{author}{Jaynes, E.~T.}
\newblock \emph{\bibinfo{title}{Probability theory: The logic of science}}
  (\bibinfo{publisher}{Cambridge university press}, \bibinfo{year}{2003}).

\bibitem{balasubramanian1997statistical}
\bibinfo{author}{Balasubramanian, V.}
\newblock \bibinfo{title}{Statistical inference, occam's razor, and statistical
  mechanics on the space of probability distributions}.
\newblock \emph{\bibinfo{journal}{Neural computation}}
  \textbf{\bibinfo{volume}{9}}, \bibinfo{pages}{349--368}
  (\bibinfo{year}{1997}).

\bibitem{zdeborova2016statistical}
\bibinfo{author}{Zdeborov{\'a}, L.} \& \bibinfo{author}{Krzakala, F.}
\newblock \bibinfo{title}{Statistical physics of inference: Thresholds and
  algorithms}.
\newblock \emph{\bibinfo{journal}{Advances in Physics}}
  \textbf{\bibinfo{volume}{65}}, \bibinfo{pages}{453--552}
  (\bibinfo{year}{2016}).

\bibitem{footnote1}
\bibinfo{note}{This difference is not our intention but result from the problem
  that $p(w \mid D^n)$ for $w \in \mathbb{R}^4$ is hard to be visualized
  without the marginalization over several parameters, such as $\Delta \omega$
  and $I_0$. For this sake, the corner plot, as described in Fig. \ref{Fig.4},
  is usually utilized to show a whole aspect of posterior probability density
  function.}

\bibitem{nagata2019bayesian}
\bibinfo{author}{Nagata, K.}, \bibinfo{author}{Muraoka, R.},
  \bibinfo{author}{Mototake, Y.-i.}, \bibinfo{author}{Sasaki, T.} \&
  \bibinfo{author}{Okada, M.}
\newblock \bibinfo{title}{Bayesian spectral deconvolution based on poisson
  distribution: Bayesian measurement and virtual measurement analytics (vma)}.
\newblock \emph{\bibinfo{journal}{Journal of the Physical Society of Japan}}
  \textbf{\bibinfo{volume}{88}}, \bibinfo{pages}{044003}
  (\bibinfo{year}{2019}).

\bibitem{tokuda2017simultaneous}
\bibinfo{author}{Tokuda, S.}, \bibinfo{author}{Nagata, K.} \&
  \bibinfo{author}{Okada, M.}
\newblock \bibinfo{title}{Simultaneous estimation of noise variance and number
  of peaks in bayesian spectral deconvolution}.
\newblock \emph{\bibinfo{journal}{Journal of the Physical Society of Japan}}
  \textbf{\bibinfo{volume}{86}}, \bibinfo{pages}{024001}
  (\bibinfo{year}{2017}).

\bibitem{inagaki2016concept}
\bibinfo{author}{Inagaki, S.} \emph{et~al.}
\newblock \bibinfo{title}{A concept of cross-ferroic plasma turbulence}.
\newblock \emph{\bibinfo{journal}{Scientific reports}}
  \textbf{\bibinfo{volume}{6}}, \bibinfo{pages}{1--6} (\bibinfo{year}{2016}).

\bibitem{yamada2008anatomy}
\bibinfo{author}{Yamada, T.} \emph{et~al.}
\newblock \bibinfo{title}{Anatomy of plasma turbulence}.
\newblock \emph{\bibinfo{journal}{Nature physics}}
  \textbf{\bibinfo{volume}{4}}, \bibinfo{pages}{721--725}
  (\bibinfo{year}{2008}).

\bibitem{tomita2017measurement}
\bibinfo{author}{Tomita, K.} \emph{et~al.}
\newblock \bibinfo{title}{Measurement of electron density and temperature using
  laser thomson scattering in panta}.
\newblock \emph{\bibinfo{journal}{Plasma and Fusion Research}}
  \textbf{\bibinfo{volume}{12}}, \bibinfo{pages}{1401018--1401018}
  (\bibinfo{year}{2017}).

\bibitem{geyer1991markov}
\bibinfo{author}{Geyer, C.~J.}
\newblock \bibinfo{title}{Markov chain monte carlo maximum likelihood}
  (\bibinfo{publisher}{Interface Foundation of North America},
  \bibinfo{year}{1991}).

\bibitem{hukushima1996exchange}
\bibinfo{author}{Hukushima, K.} \& \bibinfo{author}{Nemoto, K.}
\newblock \bibinfo{title}{Exchange monte carlo method and application to spin
  glass simulations}.
\newblock \emph{\bibinfo{journal}{Journal of the Physical Society of Japan}}
  \textbf{\bibinfo{volume}{65}}, \bibinfo{pages}{1604--1608}
  (\bibinfo{year}{1996}).

\bibitem{meng1996simulating}
\bibinfo{author}{Meng, X.-L.} \& \bibinfo{author}{Wong, W.~H.}
\newblock \bibinfo{title}{Simulating ratios of normalizing constants via a
  simple identity: a theoretical exploration}.
\newblock \emph{\bibinfo{journal}{Statistica Sinica}} \bibinfo{pages}{831--860}
  (\bibinfo{year}{1996}).

\bibitem{gelman1998simulating}
\bibinfo{author}{Gelman, A.} \& \bibinfo{author}{Meng, X.-L.}
\newblock \bibinfo{title}{Simulating normalizing constants: From importance
  sampling to bridge sampling to path sampling}.
\newblock \emph{\bibinfo{journal}{Statistical science}}
  \bibinfo{pages}{163--185} (\bibinfo{year}{1998}).

\end{thebibliography}

\section*{Acknowledgements}
This work was supported by JSPS KAKENHI (Grant number 20K19889, 20J12625, 21K03508), Shimadzu Science Foundation, and the Collaborative Research Program of Research Institute for Applied Mechanics, Kyushu University. 

\section*{Author contributions}
The work was planned and proceeded by discussion among S.T., Y.K., M.S., and S.I. S.T. carried out the Bayesian analysis. Y.K., H.A., K.Y., and K.T. conducted PANTA experiments and LIF measurements. S.T., K.Y., and M.S. finalized the manuscript with inputs from all the authors.

\section*{Competing interests}
The authors declare no competing financial interests.

\begin{table}[h]
    \centering    
    \caption{Assumed function forms of spectral intensity and corresponding velocity distribution with their parameter sets, where $\omega_{0}=448.37$ THz and $\delta=83.977$ MHz are fixed.}
    \begin{tabular}{l|l|l|l}
        \hline
        & ~~spectral intensity & velocity distribution &~~parameter set \\ \hline
        function I ~~& ~~$I(\omega; w) = I_0$~~ & no ions & ~~$w=\{I_0\}$\\
        function II ~~& ~~$I(\omega; w) = f(\omega; n_{i}, T_{i}, \Delta \omega, \omega_{0}) + I_0$~~ & Maxwellian (Eq. \eqref{eq:Maxwell}) & ~~$w=\{n_{i}, T_{i}, \Delta \omega, I_0\}$\\
        function III ~~& ~~$I(\omega; w) = \tilde{f}(\omega; n_{i}, T_{i}, \Delta \omega, \omega_{0}, \delta) + I_0$~~ & Maxwellian (Eq. \eqref{eq:Maxwell}) &  ~~$w=\{n_{i}, T_{i}, \Delta \omega, I_0\}$ \\
        function IV ~~& ~~$I(\omega; w) = g(\omega; n_{f}, \omega_b, \omega_c, \Delta \omega, \omega_{0}) + I_0$~~ & slowing-down (Eq. \eqref{eq:slowing-down}) &  ~~$w=\{n_{f}, \omega_b, \omega_c, \Delta \omega, I_0\}$ \\
        function V ~~& ~~$I(\omega; w) = \tilde{g}(\omega; n_{f}, \omega_b, \omega_c, \Delta \omega, \omega_{0}, \delta) + I_0$~~ & slowing-down (Eq. \eqref{eq:slowing-down}) &  ~~$w=\{n_{f}, \omega_b, \omega_c, \Delta \omega, I_0\}$ \\
        \hline
    \end{tabular}
    \label{tab:function_{f}orm}
\end{table}

\end{document}